# Improved Secure Efficient Delegated Private Set Intersection

Alireza Kavousi
Department of Electrical Engineering,
Sharif University of Technology,
Tehran, Iran
kavousi.alireza@ee.sharif.edu

Javad Mohajeri
Electronics Research Institute,
Sharif University of Technology,
Tehran, Iran
mohajer@sharif.edu

Mahmoud Salmasizadeh
Electronics Research Institute,
Sharif University of Technology,
Tehran, Iran
salmasi@sharif.edu

*Abstract*— **Private Set Intersection (PSI) is a vital cryptographic technique used for securely computing common data of different sets. In PSI protocols, often two parties hope to find their common set elements without needing to disclose their uncommon ones. In recent years, the cloud has been playing an influential role in PSI protocols which often need huge computational tasks. In 2017, Abadi et al. introduced a scheme named EO-PSI which uses a cloud to pass on the main computations to it and does not include any public-key operations. In EO-PSI, parties need to set up secure channels beforehand; otherwise, an attacker can easily eavesdrop on communications between honest parties and find private information. This paper presents an improved EO-PSI scheme which has the edge on the previous scheme in terms of privacy and complexity. By providing possible attacks on the prior scheme, we show the necessity of using secure channels between parties. Also, our proposed protocol is secure against passive attacks without having to have any secure channels. We measure the protocol's overhead and show that computational complexity is considerably reduced and also is fairer compared to the previous scheme.**

## I. INTRODUCTION

Secure Multi-Party Computation (SMPC) deals with problems that require distrustful parties to mutually compute a functionality without revealing their private inputs. Over the past several years, there has been growing interest among researchers to develop generic protocols that have this feature to compute any function. In this way, the main concern is efficiency. As a result, researchers shifted their focus away from generic forms and considered specific functions to achieve far better results in terms of performance.

One of the most important functionalities is the private set intersection (PSI), which enables two or more parties to find their common set inputs without any need to disclose their inputs that are not included in the intersection. Over more than a decade, PSI has been a very popular problem among researchers that also has many valuable real-world applications such as evaluating the efficiency of online advertising [16], private contact discovery [25], and location sharing [10]. Consider a situation that two security agencies want to determine their common list of defendants. Of course, the trivial solution would be revealing the list of all defendants to each other, however, due to security issues, this is certainly not the case and PSI protocol can solve this kind of problem in a secure and efficient way.

Although as we go further in years, better efficiency is obtained in PSI settings, but many of them are still nearly heavy to be used in a practical manner. One way to address this issue is to use a party, which often named a cloud server, with high capacity and computational power to take responsibility for the heavy side of the PSI computation. Computation outsourcing to a cloud alleviates some problems related to the traditional PSI, while creates some new concerns that should be regarded.

When it comes to data outsourcing, one of the biggest concerns is privacy leakage. To prevent this, owners of data prone to somehow encrypt their private data before outsourcing. Thus, this often leads to huge overhead costs. In some cases, one trick would be using blinded private data instead of encrypting them. This technique can preserve the owner's privacy against the cloud. With this regard, we can have protocols in which the cloud can process the specific operations on blinded data without any need to disclose them, and in the end, the clients by unblinding can obtain the desired results.

*A. Related work*

The first idea of PSI was presented by Meadows [2] and was almost based on the Diffie-Hellman key exchange. Many years later, the PSI problem was defined formally by Freedman et al. [4] in which their protocol was mainly constructed of additively homomorphic encryption and the notion of oblivious polynomial evaluation. Afterward, a considerable number of papers were published drawing on the idea of [4] or other cryptographic concepts such as blind-RSA and unpredictable functions [7, 8, 9]. For years, the majority of proposed protocols were based on public-key setting and therefore, this resulted in heavy computational cost and inappropriate running time. In this regard, later works such as [12, 15, 18, 20] started to take advantage of some useful techniques and tools such as balls into bins method [3], bloom filter [1], and cuckoo filter [11] to mitigate the overhead of the PSI protocols. Scalability should be considered as a crucial point in PSI protocols and when it comes to some real-world applications such as private contact discovery, it is a much more serious matter. Consequently, some papers like [22, 26] devoted their attention to unbalanced PSI in which one of the parties has limited capacity and another one has an enormous amount of it. Recently, as a result of massive complexity that parties often deal with in PSI, using a cloud as an entity to delegate heavy PSI computation to, has become a popular approach among researchers. Several schemes have been published in this context such as [14, 17, 23, 24]. Preserving privacy of party's outsourced sensitive data is an immediate concern in cloud base related schemes and there should be some considerations to prevent cloud to acquire any additional information than it is allowed. For instance, [14] is not quite private and the cloud can derive some information like cardinality of the intersection. In 2015, Abadi et al. introduced O-PSI [13] where was mainly based on additively homomorphic public-key encryption and polynomial representation of sets. [13] had many advantages over other schemes but it suffers from huge public-key complexity. Therefore, in [19], Abadi et al. presented an efficient O-PSI scheme named EO-PSI that while keeping useful features of [13], did not use any public-key operations and had far better performance than it.

*B. Our Contribution*

In [19], the authors claimed that their protocol is secure against passive attacks, but they did not state using secure channels (e.g., TLS setting) is inevitable for providing security in the scheme. In an attempt to make [19] more efficient and practical, we propose a new scheme that is secure against passive attacks, while does not need any secure channels, so the messages can be sent over public channels. Besides, our new scheme has less computational complexity compared to [19] which plays an important role in decreasing the running time of the protocol. Also, in our protocol parties' costs are more unbiased.

The rest of this paper is ordered as follows: Section II describes the required preliminaries for our scheme. In section III, an overview of EO-PSI [19] is presented. In section IV, some possible passive attacks on EO-PSI are discussed. Our new scheme is introduced in section V. Section VI represents a security analysis of the proposed scheme as well as a comparison of the proposed protocol and [19] in terms of complexity. In the end, a conclusion is given in section VII.

## II. PRELIMINARIES

In this section, we discuss some important preliminaries used in this paper. We review polynomial representation of sets and a technique called balls into bins. Also, we revisit the security model of the EO-PSI.

*A. Polynomial Representation of Sets*

Representing sets by polynomials is a widely used technique introduced in [4]. In this representation, firstly, we consider a big enough finite field $F_p$ to map all the elements of the universe set $\mathcal{U}$, then, to represent a set $S$ we can define polynomials over the field as $f(x) = \prod_{i=1}^{|S|}(x - s_i)$, where $s_i$ is a set element in $S$ and $|S| = d$. Let $f$ and $g$ be two polynomials that represent two sets $S$ and $T$ respectively, then $\gcd(f, g)$ is the polynomial describing set $S \cap T$. In [5], it is proved that if $r$ and $s$ be two degree $d$ random polynomials in $F[x]$, then $f \cdot r + g \cdot s = p \cdot \gcd(f, g)$ in which $p$ is a random polynomial in $F[x]$. As a result, we can indicate the intersection of two given sets by finding the roots of the above equation. To do so, we should factorize the equation and then consider the valid roots that corresponding to $\gcd(f, g)$ and not $p$. It is worth noting that valid roots can be distinguished from random roots by using encoding techniques in advance [5, 19].

Expressing a polynomial $f(x) = \sum_{i=0}^{d} a_i x^i$ using its coefficients($\vec{a} = [a_0, \ldots, a_d]$), causes multiplying two degree $d$ polynomials to has complexity of $O(d^2)$. A useful way to reduce this complexity to $O(d)$, is to represent a degree $d$ polynomial $f(x)$ by a set of $n$ point-value pairs $\{(x_1, y_1), \ldots, (x_n, y_n)\}$ such that $n \geq d + 1$, all $x_i$ are distinct, and $y_i = f(x_i)$ for $1 \leq i \leq n$. In this case, multiplying two degree $d$ polynomials can be done point-wisely. One thing to stress is that since $f(x) \cdot g(x)$ is a degree $2d$ polynomial, it is a need to express each polynomial at least by $n = 2d+1$ points to correctly compute the result.

*B. Balls into Bins Method*

Balls into bins method [3] is a well-known technique in computer science related problems. It is also used as a way to make PSI protocols more efficient [15,19]. Factorizing a polynomial is a highly computational task that intuitively its complexity is equal to the power of two of polynomial's degree being factorized. Therefore, it leads to an immense overhead for a $c$-elements set to be factorized when $c$ is a large number. In balls into bins technique, firstly, according to a particular hash table specification, we divide a large set to many small subsets, then execute the PSI protocol for each one. Let $H$ be a chosen cryptographic hash function mapping elements of the large set into bins numbered $1 \leq j \leq h$ as $j = H(s_i)$. The output of the hash function is a uniformly distributed number in $[1, h]$. Regarding Hash function features, the same elements mapped to the same bins. Public parameters such as hash function $H$, number of bins $h$, and maximum size of bins $d$ should be determined in advance in a way that no overflow occurs in any bins.

*C. Threat Model*

The threat model in [19] is based on the semi-honest setting in which the static adversary presents and corrupts just one of the parties at the same time. In the static semi-honest model, the corrupted party runs the assigned algorithm exactly but somehow tries to get more information about the other parties' inputs than is allowed. In this protocol, there are three parties including cloud C, client A, and client B. We consider this assumption that the cloud does not collude with clients. Here we have a functionality $F$ that should be computed alongside the protocol $\pi$ as $F: 2^{\mathcal{U}} \times \Lambda \times 2^{\mathcal{U}} \rightarrow \Lambda \times \Lambda \times 2^{\mathcal{U}}$, which three associated inputs and outputs relate to the A, C, and B respectively. $\Lambda$ Indicates the empty string, $\mathcal{U}$ indicates the whole set universe, and $2^{\mathcal{U}}$ indicates the power set of $\mathcal{U}$. To put it in another way, the functionality $F$ gets inputs just from clients A and B and then outputs the outcome just to client B. In accordance with the definition given in [6], the security of the protocol $\pi$ signifies that barely input and the corresponding output of a party is adequate for computing whatever that party can compute in the protocol. The simulation model stands for this kind of situation. Due to space constraints, we ignore the details of the simulation-based proof [6].

### III. AN OVERVIEW OF THE EO-PSI PROTOCOL

In this part, we revisit the EO-PSI protocol [19]. It should be remembered that in the EO-PSI there are three parties including cloud C, client A, and client B. The goal is to find the set intersection of A and B with the help of C. In the end, the client B finds the intersection and other parties learn nothing. This protocol includes five steps as follows.

1) **Cloud-Side Setup**

For a start, cloud C sets public parameters such as a big enough $F_p$ to map the whole universe set to, a set cardinality upper bound $c$, hash table parameters like the number of bins $h$, maximum capacity of each bin $d$, and a cryptographic hash function $H$. In addition, it chooses a vector $\vec{x}$ consists of $n$ distinct elements uniformly chosen from a finite field $F_p$, where $n = 2d+1$. Besides, the cloud chooses a pseudorandom function $PRF: \{0,1\}^m \times \{0,1\}^n \to F_p$ which pseudo-randomly maps an m-bit message to the finite field.

2) **Client-Side Setup and Data Outsourcing**

Consider $I \in \{A, B\}$ as a client who has a set $S^{(I)} \in \mathcal{U}$ and $|S^{(I)}| \leq c$. Each client does as follows:
a) Regarding to the hash table parameters, breaks down its set elements into $h$ bins randomly.
$$\forall s_i^{(I)} \in S^{(I)}: H(s_i^{(I)}) = j \quad (1)$$
where $1 \leq i \leq n$
b) Chooses a master secret key $mk^{(I)}$, then $\forall j, 1 \leq j \leq h$ generates $h$ pseudorandom values.
$$k_j^{(I)} = PRF(mk^{(I)}, j) \quad (2)$$
c) Checks every bin, if it occupies less than $d$ elements, uses random elements to increase it. Then performs the following.
  1. Represents the set corresponding to each bin by constructing a polynomial.
  $$\tau_j^{(I)}(x) = \prod_{i=1}^{d}(x - e_i^{(I)}) \quad (3)$$
  $$e_i^{(I)} = s_i^{(I)} \text{ or } e_i^{(I)} = r_{j,i}$$
  2. Computes $\tau_j^{(I)}(x_i)$ at all elements $x_i \in \vec{x}, \forall i, 1 \leq i \leq n$ to show the polynomial in point-value form.
  3. By generating pseudorandom values, blinds every value $\tau_j^{(I)}(x_i)$. To do so, uses key $k_j^{(I)}$ that was generated in the past and computes $z_{j,i}^{(I)} = PRF(k_j^{(I)}, i)$, then computes blinded values $o_{j,i}^{(I)}$ $\forall j, 1 \leq j \leq h, \forall i, 1 \leq i \leq n$ as follows:
  $$o_{j,i}^{(I)} = \tau_j^{(I)}(x_i) + z_{j,i}^{(I)} \quad (4)$$
  4. Sends $\vec{o}^{(I)} = [\vec{o}_1^{(I)}, \dots, \vec{o}_h^{(I)}]$ to the cloud.

3) **Set Intersection: Computation Delegation**
a) At first, client B sends his ID and master secret key $mk^{(B)}$ to client A.

b) Client A chooses a new key $tk$, then $\forall t, 1 \leq t \leq 3$ generates three pseudorandom values $k_t$ using this key and $\forall j, 1 \leq j \leq h$ computes pseudorandom values by each one as follows:
$$k_{t,j} = PRF(k_t, j) \quad (5)$$
c) Then, client A generates two pseudorandom polynomials $\omega_j^{(A)}(x)$ and $\omega_j^{(B)}(x)$ of degree $d$ using $k_{2,j}$ and $k_{3,j}$ respectively. Moreover, he uses $k_{1,j}$ to produce pseudorandom values for each bin as below:
$$a_{j,i} = PRF(k_{1,j}, i) \quad (6)$$
where $1 \leq i \leq n$.
d) By using keys that client A already has, he reproduces $z_{j,i}^{(A)}$ and $z_{j,i}^{(B)}$. By evaluating each of the mentioned polynomials, he computes $q_{j,i}$, $\forall i, 1 \leq i \leq n$.
$$q_{j,i} = z_{j,i}^{(A)} \cdot \omega_j^{(A)}(x_i) + z_{j,i}^{(B)} \cdot \omega_j^{(B)}(x_i) + a_{j,i} \quad (7)$$
e) Finally, client A sends $\vec{q} = [\vec{q}_1, \dots, \vec{q}_h]$ to B. Further, he sends $tk$ as well as A's and B's IDs to the cloud.

4) **Set Intersection: Cloud-Side Result Computation**
a) The cloud computes the three pseudorandom values $k_t, \forall t, 1 \leq t \leq 3$ by using $tk$ and then finds the pseudorandom values $k_{t,j}$ corresponding to each bin.
b) Cloud reproduces $a_{j,i}$, $\omega_j^{(A)}(x)$, and $\omega_j^{(B)}(x)$ using keys derived in the prior step.
c) By picking up previously outsourced datasets, the cloud computes $t_{t,j}$ as follows:
$$t_{j,i} = o_{j,i}^{(A)} \cdot \omega_j^{(A)}(x_i) + o_{j,i}^{(B)} \cdot \omega_j^{(B)}(x_i) + a_{j,i} \quad (8)$$
d) The cloud sends $\vec{t} = [\vec{t}_1, \dots, \vec{t}_h]$ to B.

5) **Set Intersection: Client-Side Result Retrieval**
a) After receiving vectors $\vec{t}$ and $\vec{q}$, client B subtracts corresponding elements of latter from former and this results in $g_{j,i}$.
$$g_{j,i} = t_{j,i} - q_{j,i}$$
$$= \omega_j^{(A)}(x_i) \cdot \tau_{j,i}^{(A)}(x_i) + \omega_j^{(B)}(x_i) \cdot \tau_{j,i}^{(B)}(x_i) \quad (9)$$
b) Client B interpolates polynomial devoted to each bin via point-value pairs $(x_i, g_i)$ and by factorizing them can reach the roots. Valid roots are regarded as the set intersection.

### IV. PASSIVE ATTACKS ON THE EO-PSI PROTOCOL

As we mentioned before, the fact that the EO-PSI protocol needs secure channels between parties to preserve privacy is not considered by the authors. In this part, we show this necessity by providing three potential passive attacks on the scheme.

1) **Obtaining $mk^{(B)}$ by an attacker in step 3.a**

In this case, according to step 2.b, the attacker can get his hands on $k_j^{(B)} = PRF(mk^{(B)}, j)$ and then regarding to step 2.c can find the pseudorandom values $z_{j,i}^{(B)} = PRF(k_j^{(B)}, i)$ for all elements in every bin. Eventually, considering outsourced dataset $o_{j,i}^{(B)} = \tau_j^{(B)}(x_i) + z_{j,i}^{(B)}$, attacker easily finds the corresponding client B's evaluated polynomials $\tau_j^{(B)}(x_i)$ and by factorizing them can learn the dataset of client B.

2) **Obtaining $q_{j,i}$ and $t_{j,i}$ by an attacker in step 3.f and 4.d**

In this case, the attacker can attain $g_{j,i}$ by subtracting $q_{j,i}$ from $t_{j,i}$ as $g_{j,i} = t_{j,i} - q_{j,i}$. Therefore, by interpolating the corresponding polynomials using point-value pairs $(x_i, g_i)$, the attacker can find the clients A and B intersection dataset.

3) **Obtaining $z_{j,i}^{(A)}$ by an attacker in step 4.d**

Regarding to the values the attacker can find during the execution of the protocol including $z_{j,i}^{(B)}$ and $tk$, he can derive other values and then by obtaining $t_{j,i}$ in step 4.d he can attain $z_{j,i}^{(A)}$. Thus, considering outsourced dataset $o_{j,i}^{(A)} = \tau_j^{(A)}(x_i) + z_{j,i}^{(A)}$, the attacker can find the dataset of client A.

Concerning the mentioned attacks, it is a need to have secure channels between parties in the EO-PSI setting. In other words, there should be some kinds of negotiating among parties in advance. Using secure channels not only manifests itself in communication and computation complexity but also in some cases may lead the protocol to be impractical.

## V. THE PROPOSED SCHEME

In this part, we propose an improved EO-PSI scheme. Similar to [19], the improved scheme involves three parties including the client C, the client A, and the client B. At the end of the protocol, client B learns the set intersection of clients A's and B's datasets. The advantage of the proposed scheme over [19] is mainly in two cases. Our new scheme does not need any secure channel to be privacy-preserving. Moreover, the total computational complexity is reduced and there is more fairness in the clients' cost than prior scheme.

1) **Cloud-Side Setup**

This step is the same as the first step of [19], and the cloud determines public parameters as before.

2) **Client-Side Setup and Data Outsourcing**

This step is the same as the second step of [19], except in the last part, blinded values $o_{j,i}^{(I)}$ are calculated as follows: $\forall j, 1 \leq j \leq h, \forall i, 1 \leq i \leq n$.

$$o_{j,i}^{(I)} = \tau_j^{(I)}(x_i) \cdot (z_{j,i}^{(I)})^{-1} \quad (10)$$

3) **Set Intersection: Computation Delegation**

a) To begin with, client B sends its ID to client A.

b) Client A recomputes the pseudorandom values $k_j^{(A)}$ using $mk^{(A)}$ as (2).

c) Client A picks a new key $tk1$, then $\forall j, 1 \leq j \leq h$ computes pseudorandom values as follows:
$$k_{tk1,j} = PRF(tk1, j) \quad (11)$$

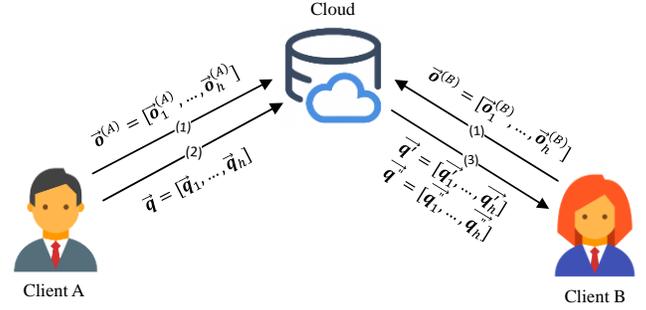

Fig. 1. Interaction between parties in the proposed scheme

d) A constructs a pseudorandom polynomial $\omega_j^{(A)}(x)$ of degree $d$ using $k_{tk1,j}$ for each bin.

e) By using keys that client A already has, he reproduces $z_{j,i}^{(A)}$. By evaluating each of the mentioned polynomials, he computes $q_{j,i}$, $\forall i, 1 \leq i \leq n$.
$$q_{j,i} = \omega_j^{(A)}(x_i) \cdot z_{j,i}^{(A)} \quad (12)$$

f) Eventually, client A sends $\vec{q} = [\vec{q}_1, \dots, \vec{q}_h]$ and IDs to the cloud C.

4) **Set Intersection: Cloud-Side Result Computation**

The cloud who has recorded the blinded outsourced datasets $\vec{o}^{(A)}$ and $\vec{o}^{(B)}$, does as follows:

a) After receiving vector $\vec{q}$, the cloud computes $q'_{j,i}$, $\forall j, 1 \leq j \leq h$, and $\forall i, 1 \leq i \leq n$ as below:
$$q'_{j,i} = q_{j,i} \cdot o_{j,i}^{(A)} = \omega_j^{(A)}(x_i) \cdot \tau_j^{(A)}(x_i) \quad (13)$$

b) By choosing a key $tk2$, the cloud C produces pseudorandom keys for each bin, $\forall j, 1 \leq j \leq h$ as follows:
$$k_{tk2,j} = PRF(tk2, j) \quad (14)$$

c) The cloud constructs $\omega_j^{(C)}(x)$ using keys derived in the prior step then computes $q''_{j,i}$, $\forall j, 1 \leq j \leq h$, and $\forall i, 1 \leq i \leq n$ in the following way.
$$q''_{j,i} = \omega_j^{(C)}(x_i) \cdot o_{j,i}^{(B)} \quad (15)$$

d) The cloud sends $\vec{q'} = [\vec{q'}_1, \dots, \vec{q'}_h]$ and $\vec{q''} = [\vec{q''}_1, \dots, \vec{q''}_h]$ to client B.

5) **Set Intersection: Client-Side Result Retrieval**

a) After receiving vectors $\vec{q'}$ and $\vec{q''}$, client B regenerates $k_j^{(B)}$, and then $z_{j,i}^{(B)}$ using the master secret key $mk^{(B)}$ and calculates $g_{j,i}$, $\forall j, 1 \leq j \leq h$, and $\forall i, 1 \leq i \leq n$.

$$g_{j,i} = q'_{j,i} + q''_{j,i} \cdot z_{j,i}^{(B)} \quad (16)$$
$$= \omega_j^{(A)}(x_i) \cdot \tau_j^{(A)}(x_i) + \omega_j^{(C)}(x_i) \cdot \tau_j^{(B)}(x_i)$$

b) Client B interpolates polynomial devoted to each bin via point-value pairs $(x_i, g_i)$ and by factorizing them can reach the roots. Valid roots are regarded as the set intersection.

**Remark 1**: Our scheme can be expanded to the multi-clients case. We do not include it in this paper for space reasons.

## VI. SECURITY AND PERFORMANCE ANALYSIS

### A. Security Analysis

We can prove that our protocol is secure according to the defined threat model and functionality $F$ in section II. In a secure PSI functionality, it should be impossible to learn anything further that is derived just by input and the corresponding output of each corrupted party during the execution of the protocol.

The proof method is similar to what is given by [19] which is based on simulation proof. Here, we ignore the concrete proof and just intuitively show resistance of our protocol against eavesdropping attacks.

1) The client B is corrupted.

According to (10) and (12), and because of not revealing the $mk^{(A)}$ and $tk1$ by client A, the adversary can only acquire the blinded values $o_{j,i}^{(A)} = \tau_j^{(A)}(x_i) \cdot (z_{j,i}^{(A)})^{-1}$ and $q_{j,i} = \omega_j^{(A)}(x_i) \cdot z_{j,i}^{(A)}$, where they do not leak any information about the dataset of the client A.

2) The client A is corrupted.

In accordance with (10), (14), and (15), the adversary can obtain $o_{j,i}^{(B)}$, $q'_{j,i}$, and $q''_{j,i}$. Thanks to adversary's inaccessibility to the master secret key $mk^{(B)}$, there is no way to compute $z_{j,i}^{(B)}$. Hence, the adversary cannot find the dataset of client B or the intersection of sets.

3) The cloud C is corrupted.

As we mentioned before, the cloud C can be corrupted by an adversary, however, it cannot collude with clients. So, in this case, due to the adversary's inaccessibility to the master secret keys $mk^{(A)}$ and $mk^{(B)}$, he cannot gain the datasets of clients A and B. Furthermore, the cloud cannot compute the (16) and learn nothing about the set intersection.

Therefore, as described above, the proposed scheme does not leak any information about datasets of client A and B or the set intersection to passive adversaries who attempt to find additional information by eavesdropping the communications between parties.

### B. Performance Analysis

In this part, we present a performance evaluation to determine the computation and communication complexities of the proposed scheme and compare the results with the EO-PSI. Identical to [19], we do not take into account the pseudorandom function cost since it can be overlooked compared with other operations cost such as modular arithmetic, interpolation, and factorization.

Tables 1, 2, and 3 indicate the computation and communication complexities of two protocols respectively. In terms of computation complexity, the proposed scheme is more efficient than [19]. As well, clients' computation complexity is more balanced. In other words, the client A's cost is reduced and the client B's cost is increased while the total complexity of the protocol is reduced. It is shown in [19] that if the maximum capacity of each bin $d$ is fixed, there is a linear relationship between the set cardinality $c$ and the number of bins $h$. It is important to note that there is no obvious difference in communication complexity between two schemes, and indeed both of the computation and communication complexities are linear with the size of the dataset.

Table 1. Computation complexity of EO-PSI [19]. We only count the online phase. Using secure channels also leads to additional cost than what is considered in this table.

| Operation | Client A | Cloud | Client B |
|---|---|---|---|
| Modular Addition | $2hn(d+1)$ | $2hn(d+1)$ | $hn$ |
| Modular Multiplication | $2hn(d+1)$ | $2hn(d+1)$ | - |
| Interpolation and Factorization | - | - | $h$ |

Table 2. Computation complexity of our scheme. We only count the online phase.

| Operation | Client A | Cloud | Client B |
|---|---|---|---|
| Modular Addition | $hnd$ | $hnd$ | $hn$ |
| Modular Multiplication | $hn(d+1)$ | $hn(d+2)$ | $hn$ |
| Interpolation and Factorization | - | - | $h$ |

Table 3. Communication complexity of EO-PSI [19] comparison with our proposed scheme.

| Protocol | Client A | Cloud | Client B |
|---|---|---|---|
| EO-PSI [19] | $O(c)$ | $O(c)$ | $O(1)$ |
| Ours | $O(c)$ | $O(c)$ | $O(1)$ |

## VII. CONCLUSION

With the growing rate of utilizing cloud computing as an important approach to solve the problem of the private set intersection, numerous studies appeared in recent years. In this paper, we scrutinized the security of an efficient delegated private set intersection EO-PSI [19] and pointed out that this protocol is vulnerable against passive attacks. We suggested a new scheme that has two advantages over the previous scheme. First, our proposed protocol is secure against eavesdropping attacks without needing any secure channels. Second, in our protocol computation complexity is reduced compared to [19] and costs are more balanced.